\documentclass{aastex}
\usepackage{spr-astr-addons}
\usepackage{url}

\RequirePackage{color}

\begin{document}

\title{Smearing of mass accretion rate variation by viscous processes in accretion disks in compact binary systems}
\shorttitle{Viscous smearing of mass accretion rate variation}
\shortauthors{Ghosh et al.}

\author{A. Ghosh} 
\affil{S. N. Bose National Centre For Basic Sciences, Kolkata}
\author{Sandip K. Chakrabarti}
\affil{ S. N. Bose National Centre for Basic Sciences, Salt Lake,
              Kolkata- 700098, India}
\affil{                    \&}
\affil{Indian Centre for Space Physics, Chalantika 43, Garia Station Road,
             Kolkata- 700084, India}



\begin{abstract}
Variation of mass supply rate from the companion can be smeared out by viscous processes inside an
accretion disk. Hence, by the time the flow reaches the inner edge, the variation in X-rays need
not reflect the true variation of the mass supply rate at the outer edge. However, if the viscosity
fluctuates around a mean value, one would expect the viscous time scale $t_{visc}$ also to spread around a
mean value. In high mass X-ray binaries, which are thought to be primarily wind-fed, the size of the
viscous Keplerian disk is smaller and thus such a spread could be lower as compared to the low mass X-ray
binaries which are primarily fed by Roche lobe overflow. If there is an increasing or decreasing 
trend in viscosity, the interval between enhanced emission would be modified systematically. 
In the absence of a detailed knowledge about the variation of mass supply rates at the outer edge,
we study ideal circumstances where modulation must take place exactly in orbital time scales, such as when
there is an ellipticity in the orbit.
We study a few compact binaries using long term All Sky monitor (ASM) data (1.5-12 keV) of Rossi X-ray
Timing Explorer (RXTE) and all sky survey data (15-50 keV) of Swift satellites by different methods to look for such smearing
effects and to infer what these results can tell us about the viscous processes
inside the respective disks. We employ three different methods
to seek imprints of periodicity on the X-ray variation and found that in all the cases, the location of the peak in
the power density spectra is consistent with the orbital frequencies. Interestingly, in
high mass X-ray binaries the peaks are sharp with high {\it rms} values, consistent with a
small Keplerian disk in a wind fed system. However, in low mass X-ray binaries with larger Keplerian disk component,
the peaks are spreaded out with much lower {\it rms} values. X-ray reflections, or superhump phenomena which may also cause such
X-ray modulations would not be affected by the size of the Keplerian disk component.
Our result thus confirms different sizes of Keplerian disks in these two important
classes of binaries. If the orbital period of any binary system  is not known, it may be obtained with reasonable accuracy for HMXBs and with lesser accuracy for LMXBs by our method.
\end{abstract}



\section{Introduction}

X-ray radiations are emitted from accretion disks around compact objects such as black holes 
primarily from two types of processes. A usually geometrically thin and optically thick disk region produces
multi-color blackbody radiation (Shakura \& Sunyaev, 1973) and a geometrically thick but generally 
optically thin component emits high energy power-law component due to repeated Compton scattering 
(Sunyaev \& Titarchuk, 1980). In a consistent model which has these two types of components built-in
as in two component advective flow (TCAF) of Chakrabarti \& Titarchuk (1995), it is understood that 
the variation in X-ray flux and spectral index is directly related to the variation of the rates 
of the two components {\it close to the black hole}. However, infall of matter is guided by the 
viscosity of the flow and had the viscosity been constant for both the components, the variation 
{\it at the outer edge} would be reflected exactly at the inner edge after a constant viscous 
time delay. Thus, for instance, if there were a periodic change in mass accretion rate and the disk size were small, a simple 
power density spectrum of a long term light curve would have revealed the periodicity with a narrow 
peak or a peak with high quality factor. But if the disk size were large, the viscosity may 
itself vary during the infall and we could expect a lesser sharp peak with a 
low quality factor. If the viscosity changes monotonically then even the period in 
X-ray emission would gradually drift.

X-ray flux from the stellar mass black holes is generally known to be time dependent. There are 
persistent X-ray sources such as Cyg X-1, Cyg X-3 etc. and there are variable sources such as GRS 1915+105
which show several classes of light curves. Yet another type of source show outbursts at irregular intervals. Examples are:
H~1743-322, GX~339-4, and GRO~J1655-40, MAXI~J1543-564 etc. Apart from strong temporal and 
spectral variations which are manifested by spectral state transitions and possibility 
of quasi-periodic oscillations, there could be subtle modulations over and above 
apparent steady and persistent flux. In order to study effects of viscous smearing, 
it is useful to target data from these time domains since viscosity could be almost 
constant and thus the there is no apparent flux or spectral state variations. Peaks 
in power density spectra could reveal orbital periodicity also.
Using two years data of RXTE/ASM for Cyg X-1, Wen et al. (1999) reported detection of the
$5.6~d$ orbital period in Lomb-Scargle periodograms of both light curves and hardness ratios when 
Cyg X-1 was in the hard state though this feature was `absent' in the soft state. 
Boroson and Vrtilek (2010) found orbital variability in Cyg X-1 during the soft 
states using light curves provided by the long-time RXTE/ASM data with a similar 
technique leading to their conclusion that the orbital variability in all soft 
states could be detected if the span of data was sufficiently longer. 
Later Wen et al. (2005) analyzed long-time (8.5 years) RXTE/ASM data using 
the same periodogram technique and showed periodic modulation in a large number 
of X-ray  sources. Their systematic analysis revealed that orbital 
modulation was more readily detected in HMXBs than in LMXBs. The fraction of eclipses from their 
observations in LMXBs was $<3\%$, which is much less compared to that of the HMXB systems. Thus, most 
surprisingly, the orbital modulation appears to be making its mark on the X-rays emitted 
from {\it the inner edge of the disk}. The information of periodicity is thus propagated through the entire Keplerian disk 
more or less faithfully independent of viscosity in the disk. 

Of course, this is not the only type of modulation seen in long term X-ray data. 
Patterson et al. (2005, and references therein) reports that cataclysmic binaries exhibit 
superhumps with periodicity a few percent higher than the orbital period provided 
the mass ratio $q=M_2/M_1<0.3$ (where, $M_2$ and $M_1$ are the companion mass and the 
compact mass respectively). Boyd, Smale and Dolan (2001) find X-ray modulation in LMC X-3 at the known orbital period
of $1.7$ days. Brocksopp et al. (1999a) find a similar modulation in Cyg X-1. Smith et al. (2002a), 
using five years of RXTE data of galactic black hole candidates 1E 1740.7-2942 and GRS 1758-258, 
show that these have periodic modulations of $12.73 \pm 0.05$ days and $18.45 \pm 0.10$ days respectively 
and interpreted these as the orbital modulations. However, Obst et al. (2013) found this 
number to be drifting for GRS 1758-258. Kudryavtsev et al. (2004) reported obtaining several 
such periodicities in the MIR satellite data and they identified some of these with periodicities 
of known objects, such as, H 1705-25, GRO J1655-40, and 4U 1543-47. They did not identify the 
reasons of such a modulation but concluded that these periodicities may not be connected to eclipse. 

A simple way to periodically modulate the inflow rate at the outer edge would be to change the 
tidal force onto the companion by the compact primary. Imagine for the sake of argument 
that there is a non-zero eccentricity of the orbit. In such an orbit, tidal force $F_t$ 
exerted by the primary on the companion would be modulated since $F_t$ is inversely proportional to the cube of 
the distance between the two stars. If the semi-major axis and the orbital eccentricity are $a$ and $e$ respectively, then 
$$
F_{t,a} \propto \frac{1}{[a(1+e)]^3}, 
\eqno{(1a)}
$$
and
$$
F_{t,p} \propto \frac{1}{[a(1-e)]^3},
\eqno{(1b)}
$$
would be the net tidal force at apoastron (marked by subscript `t, $a$') 
and periastron (marked by subscript `t,p') respectively 
(Bradt, 2008). In our context, where we consider the passage of a companion 
around a black hole, we use the terms `aponigrumcavum' and `perinigrumcavum' 
(ANC and PNC in short) to represent apoastron and periastron respectively 
(`nigrum cavum' being `black hole' in Latin). Since the supply of matter 
from the companion is proportional to this tidal effect there will be a 
periodic variation in the accretion rates, be it Keplerian flow passing 
through the Roche lobe, or just low angular momentum winds (sub-Keplerian).
Ratio of any of the component rates at ANC and PNC would be,
$$
\frac{{\dot M}_a}{{\dot M}_p} = ({\frac{1-e}{1+e}})^3 .
\eqno{(2)}
$$
Modulation of these rates will propagate in viscous time scales and would enhance 
soft X-rays which act as seed photons for Comptonization. This will, in turn, increase the flux of Comptonized X-rays 
by the same fraction. We should therefore expect a modulation of X-rays at orbital period in all 
the compact binaries whose eccentricity is non-zero and viscosity is roughly constant throughout 
the disk during the entire infall time scale.
Large amplitude variation of accretion rates in both short and long time scales is quite common and they are 
reflected in spectral properties (CT95). The enhanced matter near PNC initially in the form of 
an 'arc' is expected to be circularized due to differential motion as the perturbation 
propagates towards the black hole. Unless viscosity is very high, the shape of the density wave
is not expected to be totally smeared out (Pringle, 1981). When the viscous time scale is very small compared to the
orbital time scale, there would be multiple perturbations in the form of rings propagating inwards. However, only 
one at a time would perturb the seed photons. The next ring would affect the X-ray counts after an orbital period.
Hence, in addition to various other causes of X-ray variabilities, orbital periods would 
modulate X-rays if the orbit possesses an eccentricity.

It is well known that spectral properties of black hole candidates cannot be explained by a 
Keplerian disk alone. Along with a standard Keplerian disk, a hot electron component is 
required (Sunyaev \& Titarchuk, 1980, 1985; Zdziarski 1988; Haardt et al. 1994; Zhang et al. 2009).
In CT95 solution, this hot component is created by a low angular momentum, 
radiatively inefficient flow, which slows down close to the black hole due to centrifugal force 
and puffs up by the resulting heat. This so-called CENtrifugal pressure supported BOundary Layer, 
or CENBOL, behaves as the Compton cloud and reprocesses low energy (soft) photons 
into high energy (hard) photons. Detailed analysis of actual satellite data revealed 
that this so-called two component advective flow solution can explain even subtle aspects 
of spectral and timing properties of black hole candidates including outflow-spectral state relation and quasi-periodic oscillations
(Chakrabarti \& Manickam 2000; Rao et al. 2000; Smith et al. 2001, 2002a,b, 2007; Wu et al. 2002; Debnath et al. 2010; 
Soria et al. 2011; Nandi et al. 2012; Cambier \& Smith 2013). General success of such a model indicates
that the Compton cloud is also produced by the Companion and understanding of hard X-rays do 
not require any external source of electrons. However, a sub-Keplerian flow, arising out of companion winds 
is also expected to be modulated and this flow would arrive in almost free-fall time scale 
(if viscosity sub-critical so that it does not become a Keplerian disk). However, as CT95 found, 
that would modulate the optical depth of the `Compton cloud' and will change spectral 
slopes without changing the photon flux significantly. Smith et al. (2001, 2002b) pointed out 
that the RXTE/ASM data points to the existence of two components in the accretion flows. They find 
that there is a distinct time-lag between photon index and photon flux in low mass X-ray binary systems which 
accrete primarily through the Roche Lobe overflow, whereas high mass X-ray binaries, 
which primarily accrete winds of the companion, does not show such a lag. Thus CT95 solution naturally 
leads one to conclude that the HMXBs have small Keplerian components than the LMXBs.

In the present paper, we explore the consequence of this further. We study some systems which are known to have eccentricities, albeit small, and check if the X-ray modulation is sharp, as seen in power density spectra, for HMXBs and rather broad for LMXBs. 
Particularly we consider those systems which need not have very strong  evidence 
of eclipses of X-ray reflections from atmospheres. If the peaks are smeared out, 
with a significant error bar, this would indicate that the viscosity itself 
is highly variable. If the peaks are shifted, that could mean systematic changes in viscosity inside the disk.
We find that indeed the majority of compact X-ray binaries we study exhibit such a modulation in X-rays
with the orbital periodicity. We use publicly available RXTE/ASM and Swift Burst Alert Telescope (BAT) data from all sky survey 
instruments. We show using Fourier analysis of light curves, 
Lomb-Scargle type periodograms and e-folding light curves that the modulation at quasi-orbital period (QOP) is 
present to a varying degree. Most surprisingly, we find that for HMXBs the {\it rms} is large and for LMXBs it is small as expected. 
In future, we will consider spectral properties and how they are modified by these tidal effects. In the next Section, we present our data analysis
technique. In Section 3, we discuss known properties of several stellar mass black hole 
candidates which are relevant for our work. In Section 4, we present results of our analysis. Finally, in Section 5, we draw our conclusions.

\section{Data Analysis}

We use {\sc ascii} versions of public/archival RXTE/ASM dwell-by-dwell light curve data 
(MJD 50455 onwards) and Swift/BAT orbital light curve data (MJD 53415 onwards) 
for the X-ray binary sources Cyg X-1, Cyg X-3, XTE J1650-500, H 1705-25, 1E 1740.7-2942 and GRS 1758-258. We chose these
sources because of two reasons: (i) these are low inclination angle objects and thus the eclipsing effects would be 
negligible and (ii) most of these sources are known to have some non-zero eccentricity and thus will have a tidal  
effect mentioned above.

The RXTE/ASM has a collecting area of $90$ cm$^2$ and is operating over (1.5-12) keV range.
The Swift/BAT has an all-sky hard X-ray surveyor. It is operating over (15-150) keV
with a detecting area of $5200$ cm$^2$. We have used the data inside (15-50) keV energy range. 
In order to observe long-time behaviour of the aforesaid sources, RXTE/ASM 
data for about $13$ years and Swift/BAT data for over $8$ years without any truncation, are used. 
The standard {\sc ftools} package of {\sc heasoft} (Version 6.13) is
used for data reduction.
Neither of the instruments is meant to obtain data continuously from any particular source,
so there are `gaps' in the data, especially, due to solar constraints. 
We use a {\sc fortran} code for interpolating data at equal time intervals, required for carrying out Fourier
analysis. Gaps are padded with zeros at the same time interval. Using 
the Perl script {\sc ascii2flc}, the {\sc fits} files for the light curves 
suitable for {\sc xronos} are created. These are used to produce power 
density spectra (PDS) using {\sc powspec} task of {\sc xronos} 
(Version 5.22) package. In order to reduce noisy background and low frequency noise,
we take running average of evenly spaced data and subtract it from the initial data. 
Fluctuations of data around the running mean are then considered for producing PDS. 
Normalization factor of $-1$ is taken in {\sc powspec} task. We extracted white-noise-subtracted spectra for time
bins of $0.01$ d, $0.025$ d and $0.05$ d. Different bins were used for different objects in order to resolve
the peak in PDS well without adding noise due to low counts in each bin. 
Thus $0.05~d$ was used for Cyg X-1, GRS 1758-258 and 1E 1740.7-2942; $0.025~d$
was used for H1705-25 and $0.01~d$ was used for Cyg X-3 and XTE J1650-500.
Since observed peaks are expected to be Lorentzian type, PDS are fitted with  
Lorentzian profiles in order to determine centroids frequencies.  
The centroid frequency is considered to be due to the quasi-orbital period (QOP).
We also estimated the Q-factor (centroid frequency/width of the
Lorentzian), errors, and the significance of the observed QOPs using the {\sc fit err} 
and {\sc uncer} commands. All the QOP frequencies are estimated at 90\% confidence 
level ($1.64\sigma$). The standard {\sc qdp/plt} command, {\sc statistics}, 
is used to obtain the area under a fitted peak when zoomed above the red noise level 
(integrated rms power for a QOP), $\chi^2$ values and degrees of freedom (DOF). 
Standard conversion tables of  $\chi^2$/DOF distributions, p-values and Gaussian probabilities give 
the statistical significance of the area in terms of the standard deviation ($\sigma$) of 
Gaussian distribution. Non-constancy of viscosity and/or large Keplerian disk component is expected to reduce the sharpness of the peaks and thus
we would observe a relatively flatter peak at QOP. For convenience, we choose 
units of frequencies as $d^{-1}$ instead of $Hz$. We also plot light curves 
with actual data in modified Julian day (MJD). Note that since the overlap 
of data between the ASM and Swift/BAT is about 2000 days, results obtained from the two satellites need to be exactly the same.

As an additional check, we carried out periodogram analysis of Lomb-Scargle type from the aforesaid 
fluctuations of all the sources. This is to see if the QOPs obtained from PDS are also 
reflected generally in the periodograms. We have examined p-values and phase diagrams in all cases, 
though statistical significance in some cases were not found to be very high. We use
suitable fixed step sizes so as to get cleaner and relatively noise free peaks in the periodogram.
We further searched for the periodicities in SWIFT/BAT data using {\sc xronos} (version 5.22) 
task with command {\sc efsearch} and plotted the phase plot using the task {\sc efold}. We obtain 
similar periods. Because PDS is universally used in the literature, 
we decided to draw our conclusion about the periodicities and eccentricities 
based on the {\it rms} from PDS fits only. 

\section{X-ray Binaries under Consideration}

\subsection{Cyg X-1}

Cyg X-1 system is believed to have a stellar mass black hole and a blue supergiant star forming a binary. 
Orbital period is $5.599829\pm0.000016 d$ (Brocksopp et al. 1999b). The system does not eclipse. 
The orbital eccentricity of $0.018\pm0.003$ is very small giving rise to a nearly circular orbit
(Orosz et al. 2011). It could be appreciably high of $0.06\pm0.01$ as per another estimation (Bolton 1975).  
There is some uncertainty about the mass of the compact object and its companion. 
Stellar evolutionary models suggest a mass of $21\pm8~M_{\odot}$ turning around an O$9.7$Iab companion of 
$40\pm10~M_{\odot}$ (Zi\'{o}\l kowski 2005) while other techniques resulted in  $M_{1}=10M_{\odot}$. 
An extensive analysis of optical photometric and spectroscopic light curve data and radial velocity data along with the measurement 
of the distance between the black hole \& the companion O-star yielded a more precise value of mass 
$14.8\pm1~M_{\odot}$ and a companion mass of $M^{opt}_{2} = 19.2 \pm 1.9~M_{\odot}$ (Orosz et al. 2011). Based on a stellar evolutionary model, at the estimated distance of $2~kpc$, the companion may have a 
radius of about $15-17~R_{\odot}$ (Orosz  et al. 2011) 
and has a luminosity of approximately $3-4\times10^{5}~L_{\odot}$ (Zi\'{o}\l kowski 2005). 
The compact object is estimated to be orbiting its companion at a distance of about $40~R_{\odot}$ (Miller et al. 2005) 
and the system has an inclination $i=27.1^{\circ}\pm0.8^{\circ}$ (Orosz et al. 2011). 

\subsection{Cyg X-3}

Cyg X-3 is an accreting X-ray binary with a relativistic jet, observed from radio to high-energy gamma-rays. 
It consists of a black hole of mass $1.3-4.5~M_{\odot}$ (Zdziarski et al. 2013) which is wind-fed by a Wolf-Rayet star. 
Its distance from us is about $\sim7~kpc$. It is a high-mass X-ray binary as the companion (V1521 Cyg) is a high-mass star, having a 
mass of $7.5-14.2~M_{\odot}$. The orbital period is about $4.8~h$ (Parsignault et al. 1972; Davidsen \& Ostriker 1974)  with an
inclination of $34^{\circ}-54^{\circ}$ (Zdziarski et al. 2013). Binary separation is $d\sim3\times10^{11}~cm$. 
Only partial eclipses are observed in its otherwise obscured character at low energy. 
The companion has a strong wind ($\dot{M_{w}}\sim10^{-5}M_{\odot}~yr^{-1}$, $v_{w}\sim1000~km s^{-1}$)
(Dubus et al. 2010a). Scattering in the wind washes out rapid X-ray variability timescales 
and also modulates X-ray emission. Based on a mass-independent model of variable luminosity X-ray 
source, and an elliptic orbit, the system was reported to have an eccentricity of  $\sim0.14$ and 
a smaller orbital inclination ($i=24^{\circ}$) (Ghosh et al. 1981). However, this limit would constrain 
$M_{1}<3.6M_{\odot}$ and the companion to $M_{2}< 7.3M_{\odot}$ (Stark \& Saia 2003). 
Fermi observations show that high energy gamma-ray flux is modulated with the orbital period. 
Gamma-ray modulation is almost in anti-phase with X-ray modulation (Dubus et al. 2010b).

\subsection{XTE J1650-500}

XTE J1650-500 is a stellar mass black hole candidate probably having a mass of $3.8\pm0.5~M_{\odot}$
(Shaposhnikov \& Titarchuk 2009). It shows evidence of high frequency QPOs (Honam et al. 2003).
A safer mass range is $2.7-7.3~M_{\odot}$ (Orosz et al. 2004). Slany \& Stuchlik (2008) 
estimated the mass to be $5.1$ M$_\odot$. The binary period is $7.63~h$ and an estimation 
of the inclination is $50^{\circ}\pm3^{\circ}$ (Orosz et al. 2004). Montanari et al. (2009) suggested an evidence for an extended disk in the system. 

\subsection{H 1705-25}

It is a binary system with a small and cool star having about $0.3~M_{\odot}$ (Orosz \& Baily, 1997) 
as a companion to the black hole. The orbital period is
about $12.5~h$ (Johannsen et al. 2009). Detection of an orbital and ellipsoidal modulation 
at a period of $16.8~h$ with an orbital inclination in the range $48^{\circ}<i<51^{\circ}$ and 
distance $2-8.4~kpc$ were also reported earlier (Martin et al. 1995). 
From the characteristics of the companion star and its orbit, estimated mass of the black 
hole is $6\pm2~M_{\odot}$ (Johannsen et al. 2009).

\subsection{GRS 1758-258 and 1E 1740.7-2942}

GRS 1758-258 is often referred to as a twin source along with 1E 1740.7-2942. Both are persistent sources 
above $\sim50$ keV and are located in the vicinity of the Galactic centre.
They emit X-rays and display relativistic jets in the radio
band. The X-ray luminosity, hard spectra, persistent activity and the shape of the power spectra in  
1E 1740.7-2942 and GRS 1758-258 have made these two objects comparable to Cyg X-1 (Main et al. 1999). 
Remoteness and high column density towards GRS 1758-258 does not allow observations below (25~keV). 
Search for a counterpart in optical and infrared has not turned up any distinct result but 
only two or more candidates were found within $1''$. Weak jet structure reveals that GRS 1758-258 and 
1E 1740.7-2942 are microquasars (Keck et al. 2001).

The upper limits of the masses of the companions in GRS 1758-258 and 1E 1740.7-2942 are respectively $4M_{\odot}$ 
and  $9M_{\odot}$ with the lower limit $\sim1~M_{\odot}$ for both (Chen et al. 1994). The presence of a 
well-collimated radio jet in 1E 1740.7-2942 is an indicative of a stable accretion disk.
Companion of either of these two sources
is not able to feed the black hole via its stellar wind and they are powered by binary accretion via
Roche lobe overflow. Apparent association of the source with a high-density molecular cloud hints 
at a possible accretion directly from the ISM. The binary period in both cases is likely to be 
shorter than $20~h$ (Chen et al. 1994). Assuming that the optical component in GRS 1758-258 
is a main-sequence star and $M_1\geq3~M_{\odot}$ \& $M_2<4~M_{\odot}$, another estimation suggests 
that the period is $\leq5.8~h$ and $M_2\sim0.65~M_{\odot}$ (Kuznetsov et al. 1999). 

\section{Results of our Analysis}

In Fig. 1, we present light curves of the six sources as obtained from the ASM data 
(1.5-12 keV) of RXTE. The data duration is about 5000 days. The objects are marked. Figure 2 shows the 
power density spectra (PDS) of the reduced data obtained from the fluctuations 
around running average as discussed in Sec. 2. We also write down the names of the objects and 
time periods obtained from the Lorentzian fits. Note that the peaks are sharper in HMXB Cyg X-1 and Cyg X-3. 
In Fig. 3, we show light curves of the same objects, but drawn using all sky survey data of Swift/BAT 
(15-50 keV). Data duration is about 3000 days. The PDS extracted from the data after subtracting the dynamic mean 
are shown in Fig. 4. The time periods corresponding to QOPs are written from the Lorentzian fitting. 
In both Fig. 2 and Fig. 4, the reason for the bumps at  $\sim10^{-6}Hz$ in the PDS of Cyg X-1, 1E 1740.7-2942 \& 
GRS 1758-258, and at $\sim10^{-5}Hz$ in the PDS of Cyg X-3, XTE J1650-500 \& H 1705-25 is inherent 
to the procedure or removal of low-frequency noise by subtracting the running mean as discussed in Sec. 2.  
\begin{figure}
\includegraphics[height=8cm,width=8cm]{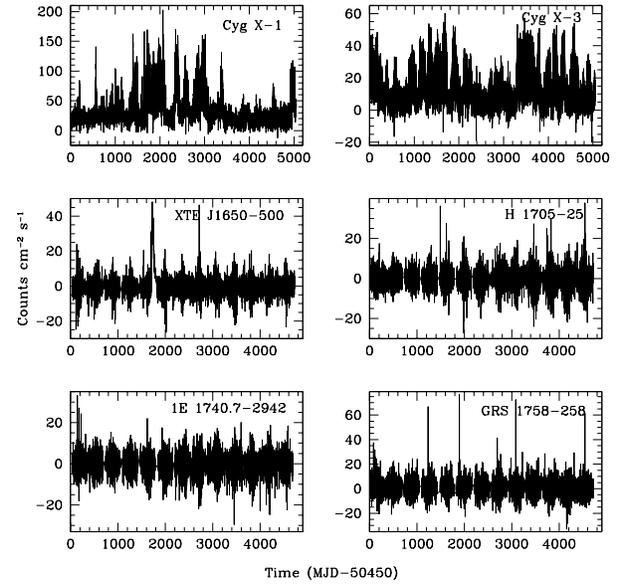}
\caption{\scriptsize Light Curves with RXTE/ASM data from MJD 50455. Energy range is (1.5 - 12) keV} 
\end {figure}

Although all the Lorentzians are fitted for $90\%$ confidence level ($1.64\sigma$), the peaks of Fig. 4 (with Swift data) 
are appreciably sharper and unique over those of Fig. 2 (with RXTE data). Interestingly, the periods obtained from 
both RXTE and Swift data agree within the error-bars. Distinct peaks at $\sim5.6~d$ and $\sim4.8~h$ are  
observed for Cyg X-1 and Cyg X-3, respectively. The prominent peak of Cyg X-3 is not due to its 
eclipsing nature since eclipsing is strong only at low energies, and not in Swift/BAT energy range, but possibly due to
smaller size of the Keplerian disk and relatively constant viscosity in the period for which Swift/BAT data was used. Modulation
of Cyg X-3 can also have a contribution due to complex interaction of the compact jets and the winds from Wolf-Rayet star (Vilhu \&
Hanninkainen, 2013). The periodicities of two enigmatic black holes GRS 1758-258 and 1E 1740.7-2942 are not known before hand.
Our analysis suggests their periodicities to be around $\sim 3.5~d$. Peaks at QOPs are particularly broad for the 
last four objects which are LMXBs which have large Keplerian disks (Smith et al. 2001, 2002b),
and thus indicative of spreading of the density perturbation region due to viscous effects (Pringle, 1981).

\begin{figure}
\includegraphics[height=8cm,width=8cm]{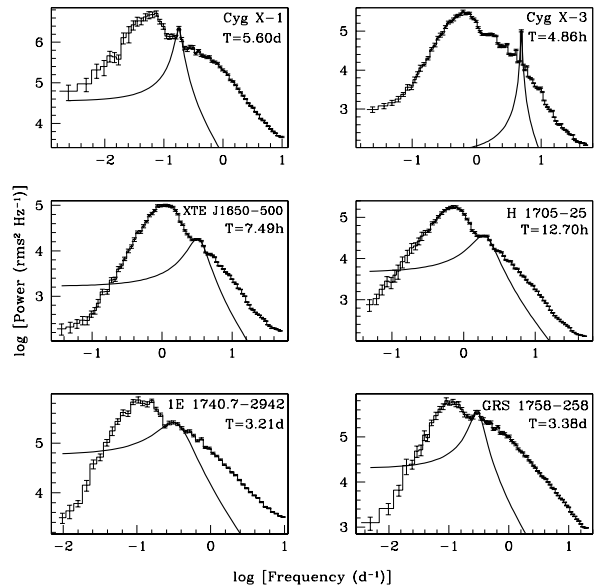}
\caption{\scriptsize Power Density Spectra with modified RXTE/ASM data. Peaks due to QOPs are fitted with Lorentzians. Centroid periods are marked in each box}
\end{figure}

We carried out an additional check on the periodicities of the obtained PDS by drawing Lomb-Scargle type periodograms 
for all these objects. In Fig. 5, we show periodograms drawn using fluctuations around the mean. The left column is 
for RXTE/ASM data and the right column is for Swift/BAT data. Though several distinct peaks are seen in 
many of these data, we see evidence of periodicities having values similar to what 
we obtained through PDS. The periods marked in each box are obtained from the strongest peaks. 
The periodicities seen in RXTE/ASM and SWIFT/BAT data are found to be the same within the error-bars.

\begin{figure}
\includegraphics[height=8cm,width=8cm]{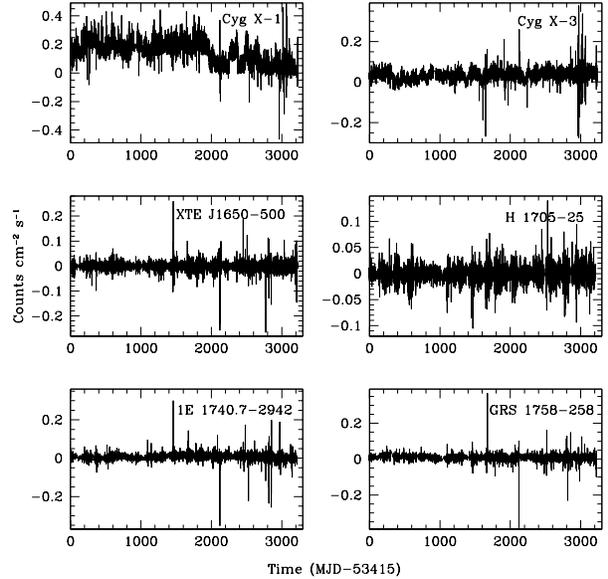}
\caption{\scriptsize Light Curves with Swift/BAT data from MJD 53415. Energy range is (15 - 50) keV}
\end{figure}

We have repeated the entire analysis using {\sc xronos} (Version 5.22) task with command {\sc efsearch}.
The periods of guess were chosen as integral multiples of $3600$s (for Cyg X-3, XTE J1650-500 \& H 
1705-25) and $86400$s (for Cyg X-1, 1E 1740.7-2942 \& GRS 1758-258) while using uneven time series 
data ({\sc ascii} version) of Swift/BAT data. Periods thus obtained have been used for 
looking into the corresponding phase behaviour using {\sc efold} command. 

\begin{figure}
\includegraphics[height=8 cm,width= 8 cm]{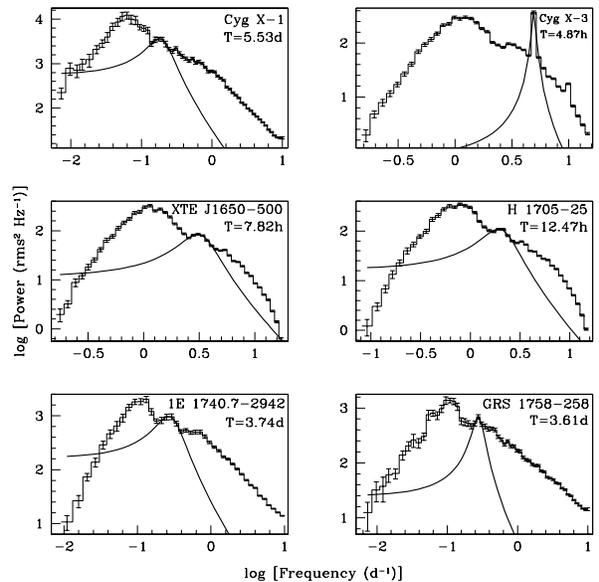}
\caption{\scriptsize Power Density Spectra with modified Swift/BAT data. Fitted QOPs are marked in each box}
\end{figure}

Table 1 summarizes results of our analysis. 
Known system parameters are given with references. In Col. 1, names of the systems are written. 
In Cols. 2 and 3, we present known masses of black holes and companions. In Cols. 4 and 5, we present inclination 
angles and periods as reported in the literature. In Cols. 6 and 7, we present centroid 
periods $T$ as obtained from power density spectra (PDS) of the complete data 
set from Swift (Col. 6) and RXTE (Col. 7) after fitting peaks with 
Lorenztian as described earlier. In Cols. 8 and 9 we present periods obtained from Lomb-Scargle periodogram 
from these two satellites. In Col. 10, we present estimated rms power (in \%).
A measure of statistical significance (in units of standard deviation $\sigma$ of Gaussian distribution) 
is given within parentheses. In Col. 11, we present the period obtained from {\sc efold} 
command. We find that the results of our analysis generally agree with observations where ever it is available. 
We also find that the objects 1E 1740.7-2942 and GRS 1758-258 may have periodicity of the order of $\sim 3.5$ days as 
opposed to reported values of $\sim 20$h  and $\sim 6$h respectively.

\section{Discussions and Conclusions}

In accreting compact binaries, X-ray flux modulations are common phenomena. There are major causes 
such as significant mass supply rate variation by the companion which leads to outbursts or class transitions. There are 
orbital modulations which are attributed to eclipsing effects, effects from wind reflections and superhumping. There could be complex
processes such as the interaction of the jets from the compact source and the winds from the Wolf-Rayet companion which may also
cause periodical changes as observed in Cyg X-3 (Vilhu \& Hannikainen, 2013). However, on the top of all these, 
there could be subtle effects which could be induced due to orbital eccentricity. In the literature,
in several compact binaries, eccentricities have been estimated (Section 3). Since that would induce 
variations in tidal forces at the apoastron and periastron, one could imagine that the resulting 
variation in accretion rates would be reflected in the light curves. These variations propagate across 
the standard Keplerian disk component in the viscous timescale, $t_{visc}$, which is given by (Frank, King \& Raine, 2002),
$$
t_{visc}=\frac{r^2}{\alpha c_s h},
$$
where, $\alpha$ is the Shakura-Sunuaev viscosity parameter ($\alpha <1$), $c_s$ is the sound speed, $h$ and $r$ are the 
height of the disk at a radial distance $r$ from the black hole. Any fluctuation in these quantities could change $t_{visc}$. 
In case $t_{visc}$ is constant (which is possible if unknown viscous stress is constant and is more likely to be the case for a small disk), 
variation of supply of matter at the outer edge would be faithfully reflected 
in the variation of seed photon flux at the inner edge (in absence of mass loss on the way) after a constant time difference. This would, in turn, modulate Comptonized photons. The peaks of resulting quasi-orbital period (QOP) would be sharp if the Keplerian disk is smaller in size and would be broader with poor quality factor if the Keplerian disk is large ($dt_{visc}/t_{visc} \sim 1$). The slope of the power density spectrum, 
especially on the low-frequency side would be separately affected by viscosity (Titarchuk \& Shaposhnikov, 2008) as well.

In the present paper, we have studied power density spectra of RXTE/ASM and Swift/BAT data for 5000 days and 3000 days
respectively for both high and low mass X-ray binaries. There was an overlap of about 2000 days in these two data sets. In all the six objects we find evidence of excess power 
at or near orbital time periods though locations of the peaks may not be exactly the same for possible non-overlap of observation dates. Our goal was not to obtain orbital periods exactly, as there are other
more accurate methods to find these orbital elements. However, our goal has been to see if smearing effects of viscosity 
can be seen in the power-density spectra. In cases of Cyg X-1 and Cyg X-3, the peaks are sharper with higher {\it rms}, indicating that 
either viscosity is almost constant and/or, what is more likely, the disk itself is very small
as was inferred by Smith et al. (2002) on the basic of explaining time lag properties using CT95 model.  
In both the Cyg X-1 and Cyg X-3, there is no signature of eclipses in our energy range. 
Partial eclipses are reported to be observed in Cyg X-3, but the system is heavily obscured only at lower energies outside 
of our range of study specially in Swift/BAT regime. If there were significant reflections from the winds, the peaks would have been much broader 
due to scattering effects and would have been independent of the size of the Keplerian disk. On the other hand, in other four objects, disk systems 
are found to be extended (Montanari et al. 2009; Smith et al. 2001, 2002b). Therefore, it is expected that even if viscosity remains 
constant there would be a considerable spread in power density spectrum at the orbital period due to turbulence. 
Of course, $t_{visc}$ being large, there is always a chance that $\alpha$ itself varies within an orbital time scale. 
This could also be seen in the case of low mass X-rays binaries since we find that the peaks of QOPs have shifted somewhat.  
Our timing analysis also predicts that the periods of 1E 1740.7-2942 and GRS 1758-258 are around $\sim 3.5$ days.

It is often believed that X-rays may be modulated by eclipsing effects of winds or reflections from the winds. However, in 
computing spectral fits, it is not usual to include effects of such eclipse or subtract effects of reflections
from the winds of the companions at certain phases. Thus these effects cannot be very important in understanding 
spectral properties. Our contention here is that the effects due to tidal forces could similarly be ignored while computing 
short time scale effects (spectra and timing properties such as phase lags, time lags and quasi-periodic oscillations), but 
in a subtle manner, tidal effects could modulate the light curves. Our present result is consistent with such a modulation in
all binary systems with non-zero eccentricity.

\begin{figure}
\includegraphics[height=15 cm, width= 12 cm]{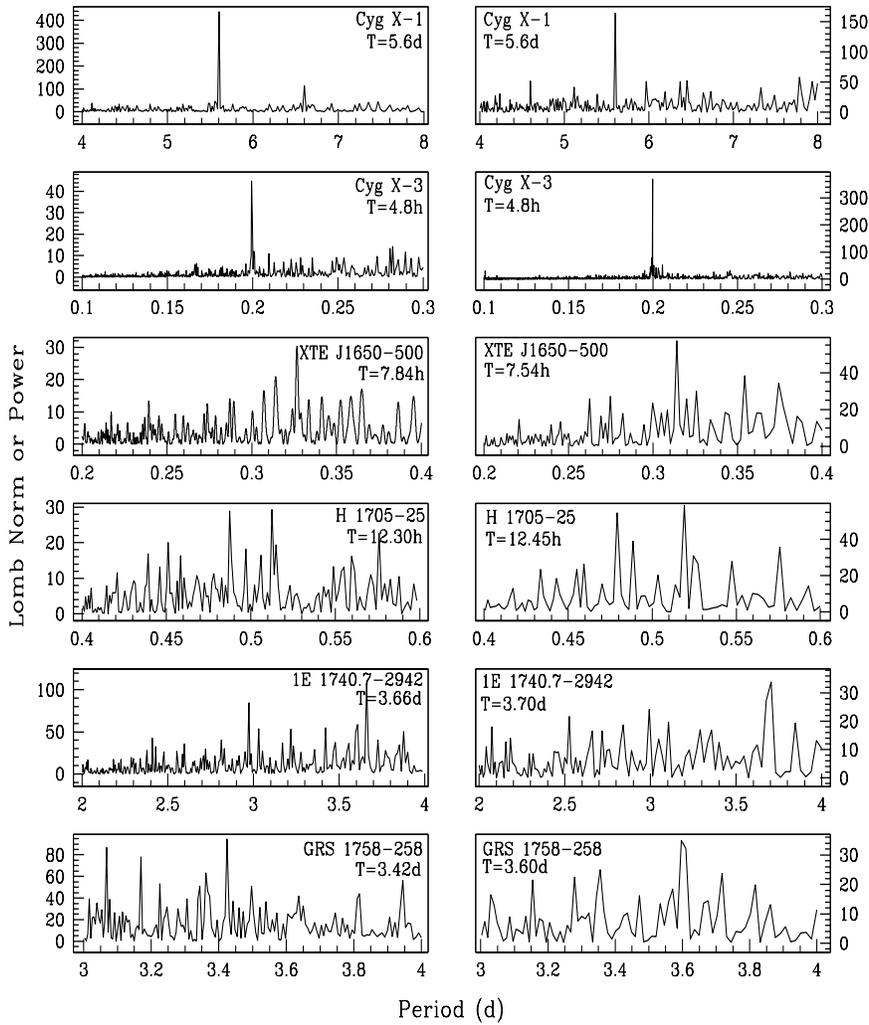}
\caption{\scriptsize Periodograms with RXTE/ASM (left) and Swift/BAT (right) data. Relevant periods are written in each box }
\end{figure}

\clearpage
\begin{table*}
\small
\caption{\bf Estimated Orbital Parameters of Six X-Ray Binaries}
\begin{tabular}{c|cccc|cccccc}
\tableline
&&&&&&\\
X-Ray Binary & $M_{1}$ &  $M_{2}$ & Incln. &  Period & \multicolumn{2}{c}{T (PDS)}&\multicolumn{2}{c}{T (LS Periodogram)}& rms (\%)& T\\
(BH/BHC)&$(M_{\odot})$&$(M_{\odot})$&$(i^{\circ})$& (T) & ({\it Swift})& ({\it RXTE})& ({\it Swift})& ({\it RXTE}) & (Stat.Sig.) & (folded)\\
\tableline
&&&&&&&&&6.72 &\\
Cyg X-1 & $14.8^{+1}_{-1}$ & $19.2^{+1.9}_{-1.9}$\tablenotemark{a} & $27.1^{+0.8}_{-0.8}$\tablenotemark{a} &  $5.6d$\tablenotemark{b} &$5.5^{+0.7}_{-0.6}d$&
$5.6^{+0.3}_{-0.3}d$ & $5.6d$ & $5.6d$  & ($3.3\sigma$) & $5.6d$\\
\tableline
&&&&&&&&&5.30 &\\
Cyg X-3 & $2.4^{+2.1}_{-1.1}$\tablenotemark{c} & $10.3^{+3.9}_{-2.8}$\tablenotemark{c} & $34-54^c$ &  $4.8h$\tablenotemark{c,d} & $4.9^{+0.2}_{-0.2}h$ & $4.9^{+0.2}_{-0.2}h$& $4.8h$& $4.8h$ & $(1.6\sigma)$ & $4.8h$\\
\tableline
&&&&&&&&&3.21 &\\
XTEJ1650-500 & $2.7-7.3$\tablenotemark{e,f} & - & $50^{+3}_{-3}$\tablenotemark{e} &  $7.63h$\tablenotemark{e} & $7.8^{+0.8}_{-0.7}h$ & $7.5^{+0.8}_{-0.7}h$&$7.5h$ & $7.8h$  &
(2.6$\sigma$) & $7.8h$\\
\tableline
&&&&&&&&&3.15 &\\
H 1705-25 & $6^{+2}_{-2}$\tablenotemark{g} & $0.3$\tablenotemark{h} &$48-51$\tablenotemark{i} & $12.5h$\tablenotemark{g} & $12.5^{+1.1}_{-0.9}h$ & $12.7^{+1.5}_{-1.2}h$ & $12.5h$ & $12.3h$ & ($1.6\sigma$) & $12.5h$\\
\tableline
&&&&&&&&&4.10 &\\
1E1740.7-2942 & - & $1-9$\tablenotemark{j} & - &  $<20h$\tablenotemark{j} & $3.7^{+0.7}_{-0.5}d$ & $ 3.2^{+0.5}_{-0.4}d$ & $3.7d$ &$3.7d$ & $(1.3\sigma)$ & $3.6d$\\
\tableline
&&&&&&&&&2.49 &\\
GRS1758-258 & $\geq3$\tablenotemark{k} & $0.65$\tablenotemark{k} & - &  $\sim6h$\tablenotemark{k} & $3.6^{+0.2}_{-0.2}d$& $3.4^{+0.3}_{-0.2}d$ & $3.6d$ & $3.4d$
& (2.0$\sigma$) & $3.6d$\\
\tableline
\end{tabular}
\tablenotetext{a}{Orosz et al. (2011)}
\tablenotetext{b}{Brocksopp et al. (1999b)}
\tablenotetext{c}{Zdziarski et al. (2013)}
\tablenotetext{d}{Davidsen \& Ostriker (1974)}
\tablenotetext{e}{Orosz et al. (2004)}
\tablenotetext{f}{Shaposhnikov \& Titarchuk (2009)}
\tablenotetext{g}{Johannsen et al. (2009)}
\tablenotetext{h}{Orosz \& Baily (1997)}
\tablenotetext{i}{Martin et al. (1995)}
\tablenotetext{j}{Chen et al. (1994)}
\tablenotetext{k}{Kuznetsov et al. (1999)}
\end{table*}


\section{Acknowledgements}
The authors thank NASA Archives for RXTE/ASM and Swift/BAT data and NASA Exoplanet Archive for producing periodograms.

\end{document}